\documentclass[aps,showpacs,twocolumn]{revtex4}
\usepackage[dvips]{graphics,graphicx}   
\newcommand{\bra}[1]{\ensuremath{\langle #1 |}} 
\newcommand{\ket}[1]{\ensuremath{| #1 \rangle}} 
\begin{document}    
\title{Multi-particle quantum chaos in tilted optical lattices}  
\author{Andrey R. Kolovsky$^{1,2}$ and Andreas Buchleitner$^1$}  
\affiliation{$^1$ Max-Planck-Institut f\"ur Physik komplexer Systeme,  
                             D-01187 Dresden}  
\affiliation{$^2$ Kirensky Institute of Physics, Ru-660036 Krasnoyarsk}  
\date{\today}    
\begin{abstract}   
We show that, in the parameter regime of state of the art experiments on Bose 
Einstein Condensates loaded into optical lattices, the energy spectrum of  
the 1D Bose-Hubbard model amended by a static field exhibits unambiguous 
signatures of quantum chaos. In the dynamics, this leads to  
the irreversible decay of Bloch oscillations.      
\end{abstract}    
\maketitle    
%%%%%%%%%%%%%%%%%%%%%%%%%%%%%%%%%%%%%%%%%%%%%%%%%%%%%%%%%%%    
Cold or ultracold atoms loaded into optical lattices (in one, two, or three 
dimensions) allow the experimental realization, with hitherto unknown 
precision, of a large variety of fascinating physical phenomena. These reach 
from quantum phase transitions \cite{Sach01},  
induced by many-particle interactions \cite{Grei02} and/or 
disorder \cite{Dam03}, over Bloch-oscillations under 
static forcing \cite{Mors01,PR}, to quantum resonances \cite{wim}, quantum 
accelerator modes \cite{arcy}, and dynamical tunneling \cite{wil}.  
Some of these phenomena are inherited from the (parameter dependent)  
properties of the ground state 
of the atoms in the extended potential, but most of them are dynamical in 
nature, and thus imply excitations of the system to energetically higher lying 
states.  
 
In the present contribution, we address the spectral properties and dynamics 
of Bose Einstein Condensates loaded into one dimensional, tilted optical 
lattices, which we model by the Bose-Hubbard Hamiltonian (with finite particle
number). The basic motivation is given by the fact that 
most state of the art experiments do not unambiguously realize the 
thermodynamic limit of large particle numbers and infinitely extended potentials,  
and that this actually is one of their decisive virtues: these systems allow, 
by systematically increasing particle number, potential extension, and, possibly, 
dimension, to monitor the {\em convergence towards} the thermodynamic 
limit. Furthermore, since these systems are {\em finite}, they exhibit a 
discrete rather than a continuous spectrum, with a density of states 
increasing with their size. Whilst increasing the density of states leads to 
an increase of the time scale over which these systems mimic the dynamics 
in the thermodynamic limit, we shall also see that their basic spectral 
properties are uncovered once the average level density is 
scaled to unity, for arbitrary systems size. Indeed, we will demonstrate that  
the spectrum of the tilted 1D Bose Hubbard problem exhibits universal Wigner-Dyson 
statistics, a hallmark of ``quantum chaos''. Such 
chaotic energy level structure implies the rapid dephasing of quantum wave 
packets, observable, e.g., in the irreversible decay of Bloch 
oscillations.  
Since irreversible wave packet dispersion, which 
ultimately means decoherence, is here brought about by many particle 
interactions, we have a case of -- experimentally tunable -- interaction 
induced decoherence.   
 
The Hamiltonian of our system reads 
\begin{eqnarray}    
\widehat{H} & = & -\frac{J}{2}\left(\sum_l    
\hat{a}^\dag_{l+1}\hat{a}_l  + h.c.\right) \nonumber \\   
 & &    
+\frac{W}{2}\sum_l\hat{n}_l(\hat{n_l}-1)+F\sum_l l\hat{n}_l \;,   
\label{10}    
\end{eqnarray}    
with $\hat{a}^\dag_{l}$,   
$\hat{a}_{l}$, and $\hat{n_l}$ the particle creation, the particle    
anihilation and the number operator at site $l$ of the lattice.  
The static field  $F$ or, more precisely, the Stark   
energy (the lattice period is set to unity), apparently destroys the 
translational symmetry of the problem. However, the latter can be restored by 
a gauge transformation which 
amounts to changing to the interaction picture with respect to the static 
field, leading to  
\begin{equation} 
\label{25} 
\widetilde{H}(t)=-\frac{J}{2}\left(e^{-i2\pi Ft}\sum_{l=1}^L  
\hat{a}^\dag_{l+1}\hat{a}_l +h.c.\right) 
\end{equation} 
\begin{displaymath} 
+\frac{W}{2}\sum_l \hat{n}_l(\hat{n_l}-1) \; , 
\end{displaymath} 
periodic in time with the Bloch period $T_{\rm B}=1/F$. Imposing periodic 
(cyclic) boundary conditions for a finite lattice size $L$, we obtain the 
quasienergy spectrum and the associated eigenvectors of $\widetilde{H}$ by 
diagonalizing the Floquet-Bloch operator  
\begin{equation} 
\label{19} 
U(T_B)=\widehat{\exp}\left(-i\int_0^{T_B} 
\widetilde{H}(t) dt\right) \;, 
\end{equation} 
where the hat over the exponential denotes time ordering. Note that, in the 
Fock basis $\{\ket{\bf n}\}$, 
defined by the $N$-particle bosonic wave function constructed from $N$  
single particle Wannier functions \cite{PR}, $U(T_B)$ 
inherits block diagonal structure from the translational symmetry of  
(\ref{25}), with each  block labeled by a suitably generalized quasimomentum 
$\kappa$ of the many-particle 
system \cite{akab03_pre}. Hence, eigenstates pertaining to different values of 
$\kappa$ do not interact, and the corresponding eigenvalues may exhibit exact 
degeneracies under variation of $F$. Therefore, to identify 
signatures of quantum chaos in the quasienergy spectrum, we have to 
unfold the spectrum such that level crossings inherited by discrete symmetries 
of the problem do not contribute, and this simply means that we have to 
restrict to sampling spectral data for fixed $\kappa$ \cite{akab03_pre}.  
Not only is this crucial for our spectral statistics, 
but it also bears very pragmatic advantages in terms of  
the required storage space in the numerical 
diagonalization of $U(T_B)$, since the dimension ${\cal J}\simeq {\cal 
  N}/L=(N+L-1)!/N!(L-1)!/L$ of a single $\kappa$ block of 
$U(T_B)$ is significantly smaller than the total dimension $\cal N$. 
 
\begin{figure}[t] 
\center 
\includegraphics[width=8.5cm, clip]{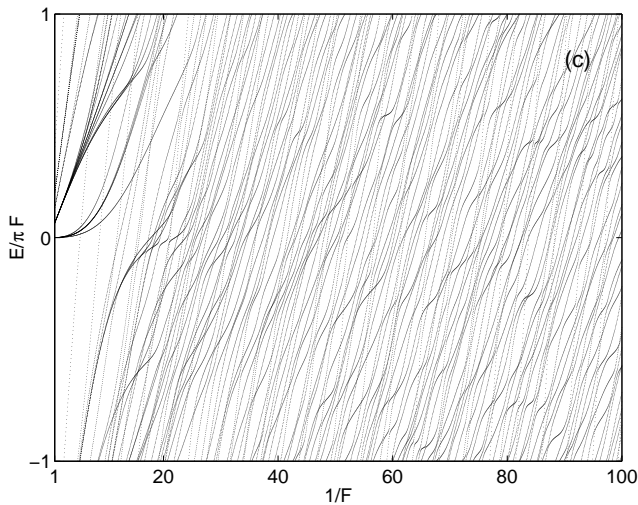} 
\caption{Evolution of the quasienergies of the Floquet-Bloch operator 
(\protect\ref{19}) under changes of the inverse static field $1/F$.  
Particle number $N=4$, lattice size $L=7$, fixed quasimomentum $\kappa =0$ 
\protect\cite{akab03_pre},  
particle-particle interaction strength $W=0.032$, hopping matrix element 
$J=0.038$. The abundance of avoided crossings of any size makes good quantum 
numbers a useless concept -- and identifies the present problem as a clear 
case of (many particle) {\em quantum chaos}.} 
\label{fig3c} 
\end{figure}

With these premises, we plot in Fig.~\ref{fig3c} the level dynamics of 
$U(T_B)$, for $J=0.038$ and $W=0.032$ (the experimental values,   
in units of photon recoil energies, for rubidium  atoms in optical lattices    
with a potential well depth of approx. ten photon recoils,  
as a function of the inverse static field 
strength. Clearly, the level structure is ``chaotic'', i.e. exhibits numerous 
avoided crossings on various scales (visible or unvisible on the scale of the 
figure), and it is rather obvious that good quantum numbers are no helpful 
concept any more (this is presicely the meaning of {\em quantum chaos}).  
Only in the strong field limit do the levels coalesce, since here the 
Stark energy largely dominates the hopping term and the particle-particle 
interaction in (\ref{10})\cite{footnote}. 
%############################################################ 
 
\begin{figure}[!t] 
\center 
\includegraphics[width=8.5cm, clip]{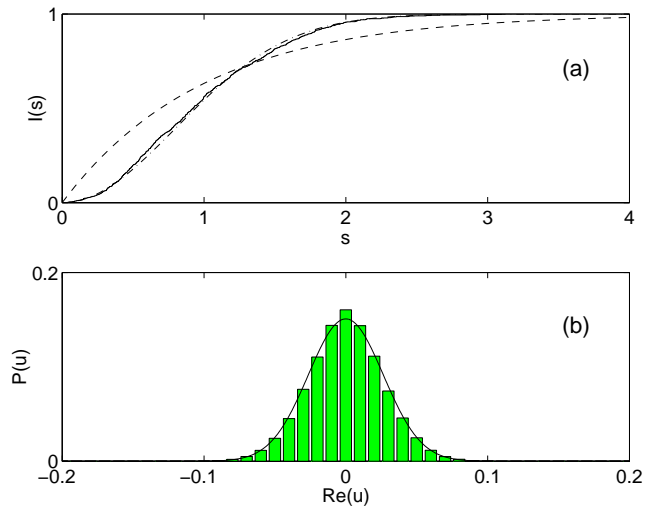} 
\caption{(a) Cumulative distribution $I(s)$ of the 
quasienergy level spacing of the Floquet-Bloch operator (full curve), 
for $F=0.01$, $W=0.032$, $J=0.038$, $N=7$, $L=9$, and fixed quasimomentum 
$\kappa$. (The dashed and dashed-dotted lines correspond  
to cumulative distributions obtained from Poissonian and Circular Orthogonal 
(COE) random 
matrix ensembles, respectively.) The numerical data unambiguously exhibit COE 
statistics, a hallmark of quantum chaos \cite{Lesh89}. 
(b) Distribution of the real parts of the matrix elements $u$ of the 
Floquet-Bloch operator, for the same parameter valus as in (a). The perfect 
fit by a Gaussian distribution underpins our identification of $U(T_B)$ with a 
random unitary matrix.} 
\label{fig7} 
\end{figure}

Fig.~\ref{fig7}(a) shows the cumulative level spacing distribution 
\begin{equation} 
I(s)=\int_0^sP(s')ds' \; , \; s=\frac{E_{j+1}-E_j}{\overline{\Delta E}}\; ,\; 
\overline{\Delta E}=\frac{2\pi F}{\cal J} 
\end{equation} 
extracted 
from a spectrum alike the one in Fig.~\ref{fig3c}, for the same values of $J$ 
and $W$, though for increased system size $N=7$, $L=9$, such as to improve the 
spectral statistics. The numerical data closely follow  
the Wigner-Dyson 
distribution for the Circular Orthogonal Ensemble (which is the appropriate 
random matrix ensemble for comparison with the unitary 
Floquet-Bloch operator \cite{Lesh89}) 
\begin{equation} 
\label{56} 
P(s)=\frac{\pi}{2}s\exp\left(-\frac{\pi}{4} s^2\right)  \;, 
\end{equation} 
and clearly exclude Poissonian statistics (associated with 
regular spectral structure). Consistently, the real (and also the imaginary)  
parts of the matrix 
elements $u=\bra{{\bf m'}}U(T_B)\ket{{\bf n'}}$ of the Floquet-Bloch operator  
are Gaussian distributed, as demonstrated by Fig.~\ref{fig7}(b), what 
underpins our identification of $U(T_B)$ with a random unitary matrix. 
We have checked that these results do neither depend on the value  
of quasimomentum, nor on the system size  
(for $N,L\le11$, $0.5<N/L<1.5$). 
Only the case $N=L$ has to be excluded, due to some additional symmetry 
property which requires a separate treatment \cite{akab03_pre}. 
%############################################################ 
 
\begin{figure}[!t]     
\center     
\includegraphics[width=8.5cm, clip]{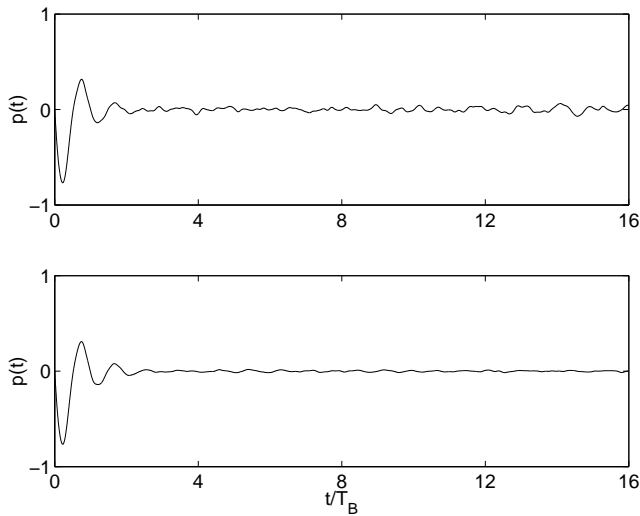}     
\caption{Irreversible decay of the Bloch oscillations of the average  
atomic momentum $p(t)$ (saled as $p\rightarrow p/NJ$) for 
a weak static field $F=0.05$, a filling factor $\bar{n}=1$ and lattice  
sizes $L=7$ (top) and $L=9$ (bottom).(The respective total   
Hilbert space dimensions are ${\cal N}=1716$ and ${\cal N}=24310$.) 
Comparison of both plots shows rapid convergence towards the  
thermodynamic limit $N,L\rightarrow\infty$, $N/L=\rm const$,  
with a well defined characteristic decay time $\tau$.}      
\label{fig4}     
\end{figure}

Let us finally address the observable consequences of the chaotic level 
structure of our many-particle Hamiltonian. The average atomic momentum  
$p(t)={\rm tr}\left[ \hat{p}\hat{\rho}(t)\right]$ is a quantity readily 
accessible in state of the art experiments \cite{Mors01,Daha96}, and 
determined by the single particle density   
matrix $\hat{\rho}(t)$, which can be easily deduced from the total wave 
function $\ket{\Psi(t)}$ generated by the action of  
$U$ on the initial condition $\ket{\Psi(0)}$. In the absence of 
particle-particle interactions, $p(t)$ exhibits the well-known Bloch 
oscillations \cite{PR} with Bloch frequency $\omega_B=2\pi F$. For 
non-vanishing $W$, however, the off-diagonal elements of $\hat{\rho}(t)$ decay 
on a characteristic time scale $\tau$, leading to {\em irreversible} 
decay of the Bloch oscillations, as born out in Fig.~\ref{fig4} for different 
system sizes, at fixed filling factor $N/L=1$ of the lattice. 
%################################################################ 
Note that the time dependence of $p(t)$ is virtually identical for both 
choices of $N$ and $L$ in Fig.~\ref{fig4}, for times shorter than 
approx. three Bloch periods. Only for longer times do residual fluctuations 
persist, which, however, are quickly damped out with increasing Hilbert space 
dimension. Hence, at least as far as this specific dynamical observable is 
concerned, the thermodynamic limit is rapidly approached, simply due to the 
rapid increase of Hilbert space -- and, hence, of the level density -- with $N$ 
and $L$. Whilst the decay of the wave packet is clearly due to the chaotic 
level structure of Fig.~\ref{fig3c} and is brought about by the rapid dephasing of 
the various spectral components which constitute the initial $\ket{\Psi(0)}$, 
we have so far no quantitative understanding of the dependence of $\tau$ on 
$W$, $J$, and $F$. Clearly, this is a challenging and equally exiting problem, 
since experimentalists will be able to enter the regime of 
multiparticle quantum chaos in a controlled way, and hence to test the genesis 
of interaction induced decoherence in the strictly unitary dynamics ruled by 
the tilted Bose-Hubbard Hamiltonian.

 %%%%%%%%%%%%%%%%%%%%%%%%%%%%%%%%%%%%%%%%%%%%%%%%%%%%%%%%%%%%%%%% 
     

\begin{thebibliography}{10}  
\bibitem{Sach01}     
S.~Sachdev, Quantum phase transitions (Cambridge Univ. Press., Cambridge, 2001). 

\bibitem{Grei02}     
M.~Greiner et al.,   
%O. Mandel, T.~Esslinger, T.~W.H\"ansch, and I.~Bloch,     
Nature \textbf{415}, 39 (2002).  

\bibitem{Dam03} 
B. Damski et al., Phys. Rev. Lett. {\bf 91}, 080403 (2003). 

\bibitem{Mors01} 
O. Morsch et al., Phys. Rev. Lett. {\bf 87}, 140402 (2001). 

\bibitem{PR}     
M.~Gl\"uck, A.~R.~Kolovsky, and H.~J.~Korsch,      
Phys. Rep. \textbf{366}, 103 (2002).    
  
\bibitem{wim} 
S. Wimberger, S. Fishman, and I. Guarneri, Nonlinearity {\bf 16}, 1381 (2003). 

\bibitem{arcy} 
S. Schlunk et al., Phys. Rev. Lett. {\bf 90}, 124102 (2003).  

\bibitem{wil} 
W.K. Hensinger et al., Nature {\bf 412}, 52 (2001).
  
\bibitem{akab03_pre}  
A. Kolovsky and A. Buchleitner, Phys. Rev. E, in print (2003). 
     
\bibitem{footnote} In this case,  
the Floquet-Bloch operator is diagonal in the Fock basis representation  

\protect\cite{akab03_pre}, 
leading to decay and revival of Bloch oscillations across the lattice, with a 
interaction induced revival time $\sim 1/W$ (also see    
A.~R.~Kolovsky, Phys. Rev. Lett. {\bf 90}, 213002 (2003)).   
  
\bibitem{Lesh89}     
Chaos and quantum physics, eds. M.-J.~Giannoni, A.~Voros,     
and J.~Zinn-Justin (North-Holland, Amsterdam, 1991).     
 
\bibitem{Daha96}     
BenDahan et al., Phys. Rev. Lett. \textbf{76}, 4508 (1996).  
\end{thebibliography}
\end{document}